\documentclass[preprint,showpacs,preprintnumbers,amsmath,amssymb]{revtex4}

\usepackage{graphicx}
\usepackage{dcolumn}
\usepackage{bm}

\begin{document}


\title{Clustering features of the $^7$Be nucleus in relativistic fragmentation}

\author{N.~K.~Kornegrutsa}
   \affiliation{Joint Insitute for Nuclear Research, Dubna, Russia}
 \author{D.~A.~Artemenkov}
   \affiliation{Joint Insitute for Nuclear Research, Dubna, Russia} 
  \author{R.~R.~Kattabekov}
   \affiliation{Institute for Physics and Technology, Uzbek Academy of Sciences, Tashkent, Republic of Uzbekistan} 
 \author{K.~Z.~Mamatkulov}
   \affiliation{Djizak State Pedagogical Institute, Djizak, Republic of Uzbekistan}    
 \author{P.~I.~Zarubin}
     \email{zarubin@lhe.jinr.ru}    
     \homepage{http://becquerel.jinr.ru}
   \affiliation{Joint Insitute for Nuclear Research, Dubna, Russia} 
 \author{I.~G.~Zarubina}
   \affiliation{Joint Insitute for Nuclear Research, Dubna, Russia}   

\date{\today}

\begin{abstract}
\indent  Charge topology of fragmentation of 1.2 A GeV $^7$Be nuclei in nuclear track emulsion is presented. The dissociation channels $^4$He + $^3$He,
 2$^3$He+ n, $^4$He + 2$^1$H are considered in detail. It is established that the events $^6$Be + n amount about to 27 \% in the channel $^4$He + 2$^1$H.

\end{abstract}

\maketitle
\section{}
\indent Stacks of pellicles of nuclear track emulsion provide a special opportunity to explore clustering of light nuclei
 (reviewed in \cite{Zarubin}).
 The presented results on dissociation of $^7$Be nuclei are demonstrate the  progress in research carried out by the BECQUEREL Collaboration.
 The $^7$Be nucleus
 is a source for the study of the states $^3$He + $^4$He, $^3$He + $^3$He + n, $^6$Li + p and $^6$Be + n. The pattern of fragmentation is
 important for understanding of the structure features of the nuclei $^8$B, $^9$C and $^{12}$N because the $^7$Be nucleus plays the role 
 of a core in them.\par
 \indent Nuclear track emulsion was irradiated at the Nuclotron of the Joint Institute for Nuclear Research (JINR, Dubna) by a mixed beam 
 of $^7$Be, $^{10}$C,
 and $^{12}$N nuclei which was created by selecting products of charge-exchange and fragmentation processes involving $^{12}$C nuclei accelerated to an
 energy of 1.2 GeV per nucleon \cite{Rukoyatkin}, \cite{Kattabekov10}, \cite{Mamatkulov}, \cite{Kattabekov13}. Viewing of the exposed 
 pellicles and the track classification made it possible to establish the charge topology of
 the $^7$Be nucleus. Peripheral fragmentation distribution of the 289 found events N$_{ws}$ not accompanied by target fragments 
 (\lq\lq white \rq\rq stars)
 is presented in Table 1 over the fragmentation channels as well as 380 events N$_{tf}$ accompanied by target fragments.\par
 \indent The distribution of the 79 events 2He which were successfully identified by multiple scattering is presented in Table 2. It gives an idea about
 the relationship configurations $^3$He + $^4$He and 2$^3$He + n in the $^7$Be structure, as the identification was carried out without bias. The channel
 $^3$He + $^4$He dominates over 2$^3$He indicating on a higher probability of the two-body configuration $^3$He + $^4$He in the $^7$Be structure compared
 to 2$^3$He + n. The probability of the 2$^3$He + n channel is significant, amounting to about 30\%.\par
\begin{figure}
\centering
  \includegraphics[width=5in]{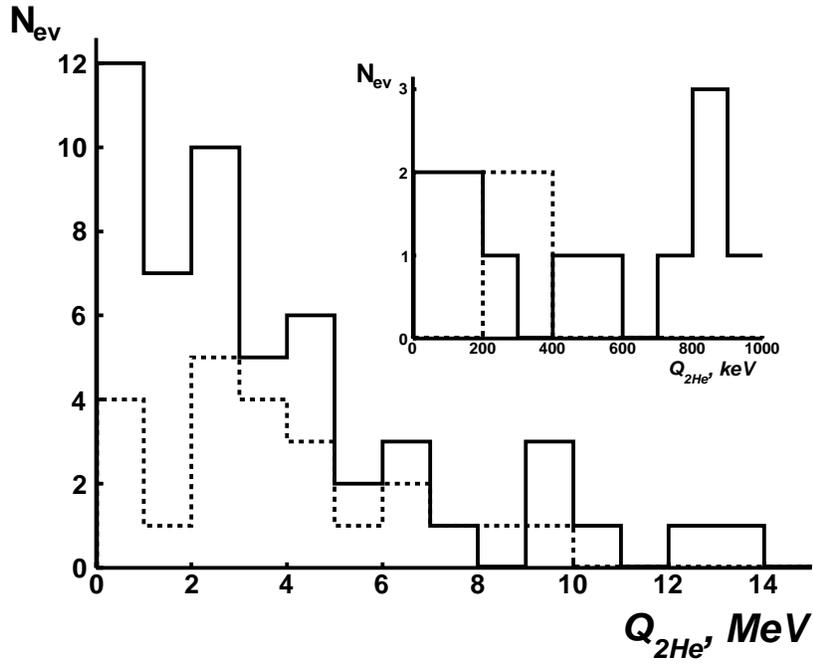}
\caption{\label{fig:1}  Distribution of events Distribution of events $^7$Be$\rightarrow ^3$He + $^4$He and 2$^3$He over the excitation energy Q$_{2He}$
 (solid and dotted histograms, respectively). Histograms for values Q$_{2He} < $1 MeV are on the insertion and 2$^3$He over the excitation
 energy Q$_{2He}$ (solid and dotted histograms, respectively). Histograms for values Q$_{2He} < $1 MeV are on the insertion.}
\end{figure}

 \indent The distribution of the events 2$^3$He and $^3$He + $^4$He over the excitation energy Q$_{2He}$ defined as a difference of the invariant mass of
 the fragmenting system and the sum of the fragment masses is shown in Fig.~\ref{fig:1}. The values Q$_{2H}$ of the channel $^3$He + $^4$He
 are distributed in the
 range covering the known levels of the $^7$Be nucleus excitation.\par
 \indent One of the tasks of this study consisted in searching for narrow pairs 2$^3$He with values Q$_{2^3He}$ in a range of 100 -- 200 keV 
 the indication
 to which was obtained for dissociation $^9$C$\rightarrow$3$^3$He. The obtained distribution includes four events with values in the range of
 200--400 keV (Fig.~\ref{fig:1}, dotted histogram in insertion). These data do not exclude a possible existence of the resonant state 2$^3$He discussed in \cite{Artemenkov}.\par
\begin{figure}
\centering
  \includegraphics[width=5in]{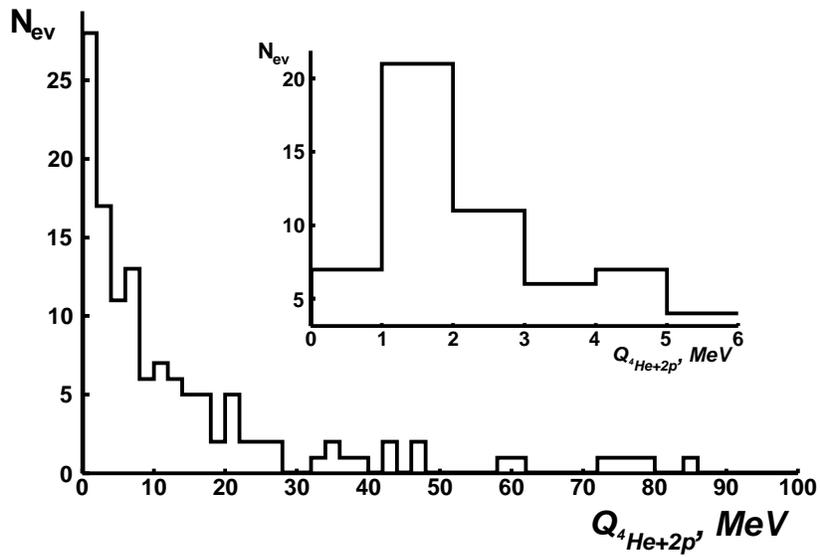}
\caption{\label{fig:2} Distribution of events $^7$Be$\rightarrow ^4$He + 2p over the excitation energy Q$_{^4He+2p}$.}
\end{figure}

 \indent There is an opportunity of $^7$Be fragmentation via an unstable $^6$Be nucleus with a threshold 1.37 MeV above $^4$He + 2p. Fig.~\ref{fig:2} shows
 distribution of events $^4$He + 2p over the difference of the invariant mass of the produced $\alpha$-particle and two protons and their mass sum
 Q$_{^4He+2p}$. The region Q$_{^4He+2p} <$ 6 MeV indicates on the presence of about 27\% events $^7$Be $\rightarrow ^6$Be $\rightarrow ^4$He + 2p. Thus,
 contribution of the configuration $^6$Be + n to the $^7$Be structure is estimated at a level of 8 $\pm$ 1 \%. \par
\begin{figure}
\centering
  \includegraphics[width=5in]{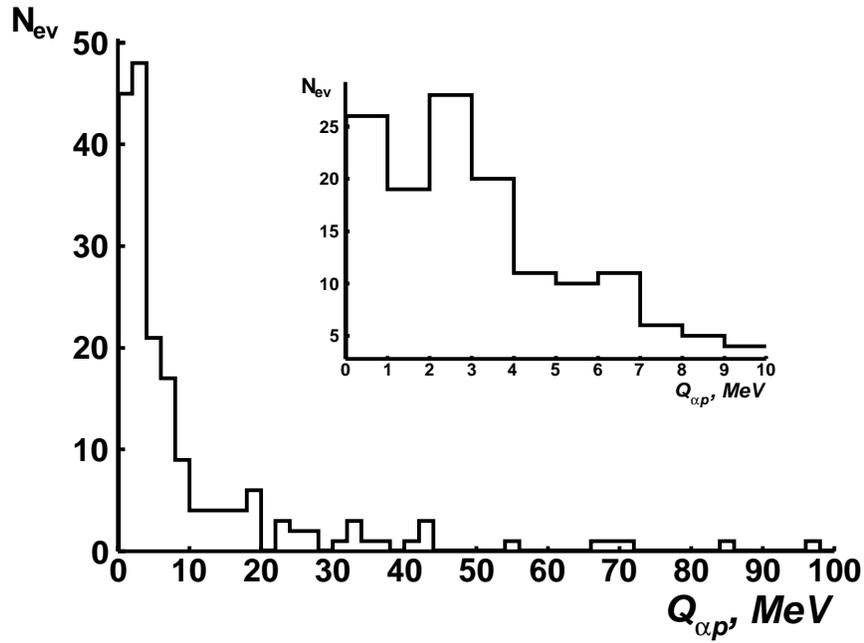}
\caption{\label{fig:3} Distribution of events $^7$Be$\rightarrow ^4$He + 2p over the excitation energy Q$_{^4He+p}$ (events related to the $^6$Be decays are excluded
 from this histogram.)}
\end{figure}

 \indent The question about the contribution of the $^5$Li resonance decaying to $\alpha$ + p with an energy of 1.69 MeV and width of 1.5 MeV has a 
 significance independent of $^6$Be since the production threshold of $^5$Li + p is 0.35 MeV higher than the one of the ground state $^6$Be.
 Despite of the absence of a clear signal the distribution Q$_{\alpha p}$ (Fig.~\ref{fig:3}) does not contradict to a possible contribution of $^5$Li decays.\par
 \indent This work was supported in part by the Russian Foundation for Basic Research (project no. 12-02-00067) and by grants from the plenipotentiaries
 of Bulgaria and Romania at the Joint Institute for Nuclear Research (Dubna).\par
 \begin{table}[t]
\caption{Distribution over the dissociation channels of $^7$Be nuclei for \lq\lq white\rq\rq stars
 N$_{ws}$ and events with target fragments or produced mesons N$_{tf}$.}
\centering
\label{tab:1}       
\begin{tabular}{lcccc}
\hline\noalign{\smallskip}
Channel & 2He & He + 2H & 4H & Li + H\\[3pt]
\hline\noalign{\smallskip}
N$_{ws}$ & 115 & 157 & 14 & 3 \\
N$_{tf}$ & 154 & 226 & - & -\\
\hline\noalign{\smallskip}
\end{tabular}
\end{table}
\begin{table}[t]
\caption{Distribution over the dissociation channels of $^7$Be nuclei for \lq\lq white\rq\rq stars
 N$_{ws}$ and events with target fragments or produced mesons N$_{tf}$.}
\centering
\label{tab:2}       
\begin{tabular}{lcc}
\hline\noalign{\smallskip}
Channel & $^3$He + $^4$He &   $^3$He + $^3$He\\[3pt]
\hline\noalign{\smallskip}
N$_{ws}$ & 32 & 14 \\
N$_{tf}$ & 24 & 9 \\
\hline\noalign{\smallskip}
\end{tabular}
\end{table}



\newpage


\begin{thebibliography}{}
\bibitem{Zarubin}
 Zarubin P. I., Lect. Notes in Phys., 875, Clusters in Nuclei, Vol. 3. Springer International Publishing, pp 51-93. 2013; arXiv:1309.4881. 
\bibitem{Rukoyatkin}
 Rukoyatkin P. A. et al., EPJ ST 162, 267 (2008); arXiv:1210.1540.
\bibitem{Kattabekov10}
 Kattabekov R. R. et al., Phys. Atom. Nucl. 73, 2110 (2010); arXiv:1104.5320.
\bibitem{Mamatkulov} 
 Mamatkulov K. Z. et al.  Phys. At. Nucl., 2013, Vol. 76, No. 10, pp. 1224-1229; arXiv:1309.4241.
\bibitem{Kattabekov13}
 Kattabekov R. R.  et al. Phys. At. Nucl., 2013, Vol. 76, No. 10, pp. 1219?1223; arXiv:1310.2080.
\bibitem{Artemenkov} 
 Artemenkov D. A.  et al., J. Phys. Conf. Ser. 337 (2012) 012019 arXiv:1105.3813.
 
\end{thebibliography}
\end{document}